\documentclass[3p,times]{elsarticle}

\usepackage{ecrc}
\volume{00}
\firstpage{1}
\journalname{Nuclear Physics A}
\runauth{A. Buzzatti and M. Gyulassy}
\jid{nupha}

\usepackage{amssymb}
\usepackage[figuresright]{rotating}
\usepackage{wrapfig}
\usepackage{epsfig}
\usepackage{dcolumn}
\usepackage{bm}
\usepackage{color}

\newcommand{\be}{\begin{equation}}
\newcommand{\ee}{\end{equation}}
\newcommand{\bea}{\begin{eqnarray}}
\newcommand{\eea}{\end{eqnarray}}

\newcommand{\bx}{\mathbf{x}}

\begin{document}

\begin{frontmatter}
\title{An overview of the CUJET model: \\ Jet Flavor Tomography applied at RHIC and LHC}
\author{Alessandro Buzzatti}
\ead{buzzatti@phys.columbia.edu}
\author{Miklos Gyulassy}
\address{Department of Physics, Columbia University, 538 West 120th Street, New
York, NY 10027, USA}
\begin{abstract}
Jet Flavor Tomography is a powerful tool used to probe the properties of Quark
Gluon Plasma formed in heavy ion collisions at RHIC and LHC.
A new Monte Carlo model of jet quenching developed at Columbia University,
CUJET, was applied to predict the jet flavor and centrality dependence of the
nuclear modification factor $R_{AA}$.
The predictions for fragments $f=\pi,D,B,e$, derived from quenched jet flavors
$a=g,u,c,b$ in central and peripheral collisions at RHIC and LHC, exhibit novel
features such as a level crossing pattern in $R_{AA\rightarrow a\rightarrow f}$
over a broad transverse momentum range which can test jet-medium dynamics in
quark gluon plasmas and help discriminate between current energy loss models.
Furthermore, the inclusion of running coupling effects seems to change the jet
energy dependence of the jet energy loss to a non trivial constant behavior,
with a visible impact on the predictions for $R_{AA}$.
\end{abstract}
\end{frontmatter}

\section{Introduction}

We report results of a new Monte Carlo pQCD tomographic model, CUJET
\cite{CUJET},
developed as part of the of the ongoing Department of Energy JET Topical
Collaboration \cite{JETColl} effort to construct more powerful numerical codes,
which are
necessary to reduce the large theoretical and numerical systematic uncertainties
that have hindered so far quantitative jet tomography.
CUJET extends the development of the the GLV, DGLV and WHDG \cite{GLV,DGLV,WHDG}
opacity
series approaches by including several dynamical features such as:\\
(1)	
dynamical jet interaction potentials that can interpolate between the pure HTL
dynamically screened magnetic \cite{MD} and static electric screening
\cite{GLV,DGLV} limits;\\
(2)	
the ability to calculate high order opacity corrections to interpolate
numerically between $N=1$ and $N=\infty$ analytic approximations;\\
(3)	
full jet path proper time integration over longitudinally expanding and
transverse diffuse QGP geometries;\\
(4)	
the ability to evaluate systematic theoretical uncertainties such as sensitivity
to formation and decoupling phases of the QGP evolution, local running coupling
and screening scale variations, and other effects out of reach with analytic
approximations;\\
(5) 
elastic in addition to radiative fluctuating energy loss distributions;\\
(6) 
convolution over $\sqrt{s}$ and flavor dependent pQCD invariant jet spectral
density (without local in $p_T$ spectral index approximations);\\
(7) 
convolution over final fragmentation, $D_{f/a}(x,Q)$, as well as semileptonic
decay distributions.

The model is here applied to predict the nuclear modification factor
$R_{AA\rightarrow a\rightarrow f}(y\approx 0, p_T, \sqrt{s} ,{\cal C}=0-5\%)$,
for a variety of jet parton flavors $a=g,u,c,b$ and final fragments
$f=\pi,D,B,e^-$, over a broad $p_T$ kinematic range at mid-rapidity for central
collisions at $\sqrt{s}=0.2,2.76$ GeV as observed at RHIC \cite{RHIC} and LHC
\cite{LHC}.
Our attempt is to provide an improved and detailed implementation of the opacity
series based theoretical models and consequently set more stringent constraints
on the nature of the Quark Gluon Plasma and its coupling with highly energetic
jets.

\section{CUJET model}
We remind that CUJET uses Monte Carlo techniques to compute finite order in
opacity contributions to the jet medium induced gluon radiative spectrum.
In the following analysis, we limit our studies to the first order in opacity;
furthermore, we model the interaction potential between the jet and the medium
with the pure HTL dynamical model.
Neither of these approximations are shown to qualitatively alter the results.
The energy loss is computed starting from the induced radiated gluon number per
collinear light cone momentum fraction, $dN/dx_+$, where fluctuations of the
radiated gluon number are included via Poisson expansion (incoherent emission).
Elastic energy losses, estimated using the Thoma-Gyulassy model, are
accounted for as well, and show a contribution close to $25-30\%$ of the total
energy loss.

The plasma is assumed inhomogeneous (Glauber profile) and nonstatic (1+1D
Bjorken expansion), and its density profile is constrained solely by the initial
observed rapidity density $dN/dy$, equal here to $1000$ for $\sqrt{s}=0.2$ GeV
RHIC collisions and $2200$ for $\sqrt{s}=2.76$ GeV LHC collisions; we also
set the initial thermalization time $\tau_0=1$ fm/c and take $T\approx 100$ MeV
as the characteristic temperature at which the jet decouples from the medium.

After obtaining the relative energy loss ($\epsilon$) probability distribution functions
$P_a(\epsilon; p_i, \bx,\phi)$ for a parton $a=g,u,c,b$ of given initial
momentum $p_i$, the average over initial transverse jet production points $\bx$
and directions $\phi$ is taken according to the standard binary collision
density and convoluted over the pQCD initial jet flavor invariant $pp$ cross
sections
\be
\label{Raa}
	\frac{d\sigma_a(p_f)}{dyd^2 p_f} \equiv
 R_{AA}^a(p_f) \frac{d\sigma^0_a(p_f)}{dyd^2 p_f} \nonumber 
=\left\langle \int 
d\epsilon\; P_a(\epsilon; p_i ,\bx,\phi )  
\left(\frac{d^2p_i}{d^2p_f}\right) 
\frac{d\sigma_a^0(p_i)}{dyd^2 p_i}\right\rangle_{\bx,\phi} 
 \;\;.
\ee

Finally, we assume in-vacuum fragmentation and compute the observed final state
nuclear modification factors.

\section{Results}

\begin{figure*}[tbh]
\label{Fig1}
\includegraphics[width=3.in
]{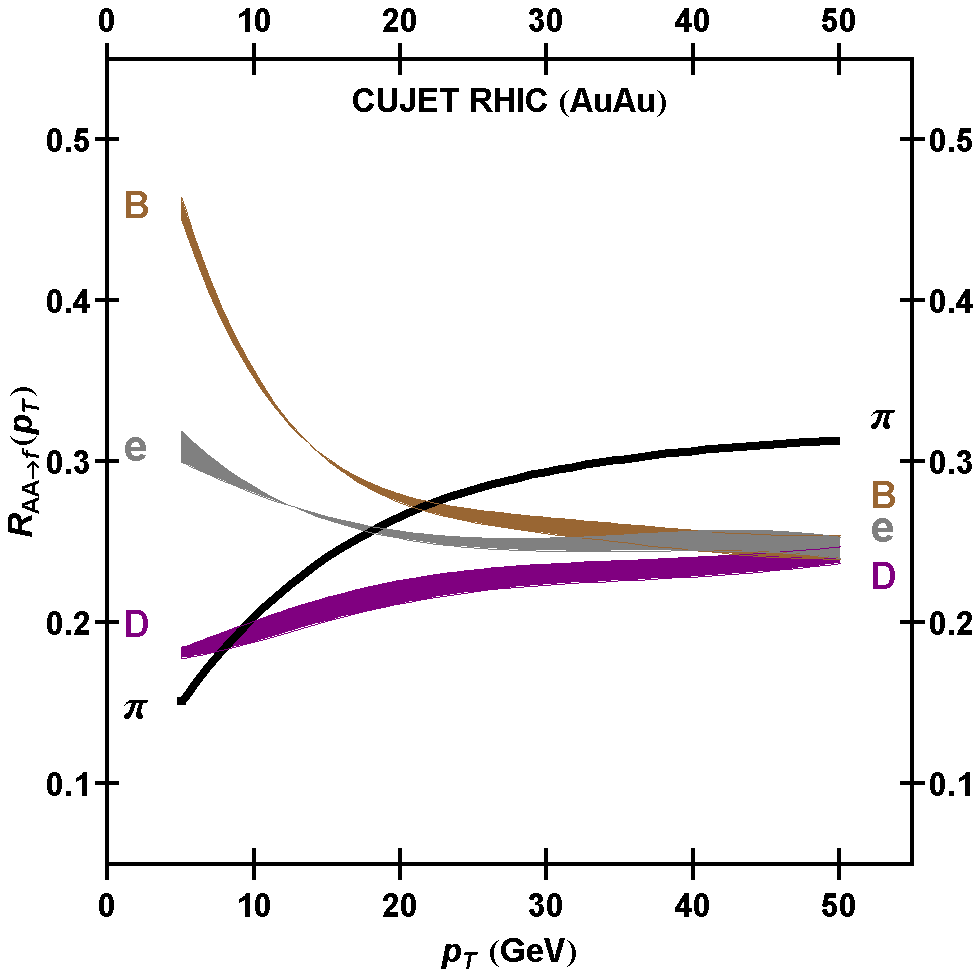}
\hspace{0.25in}
\includegraphics[width=3.in
]{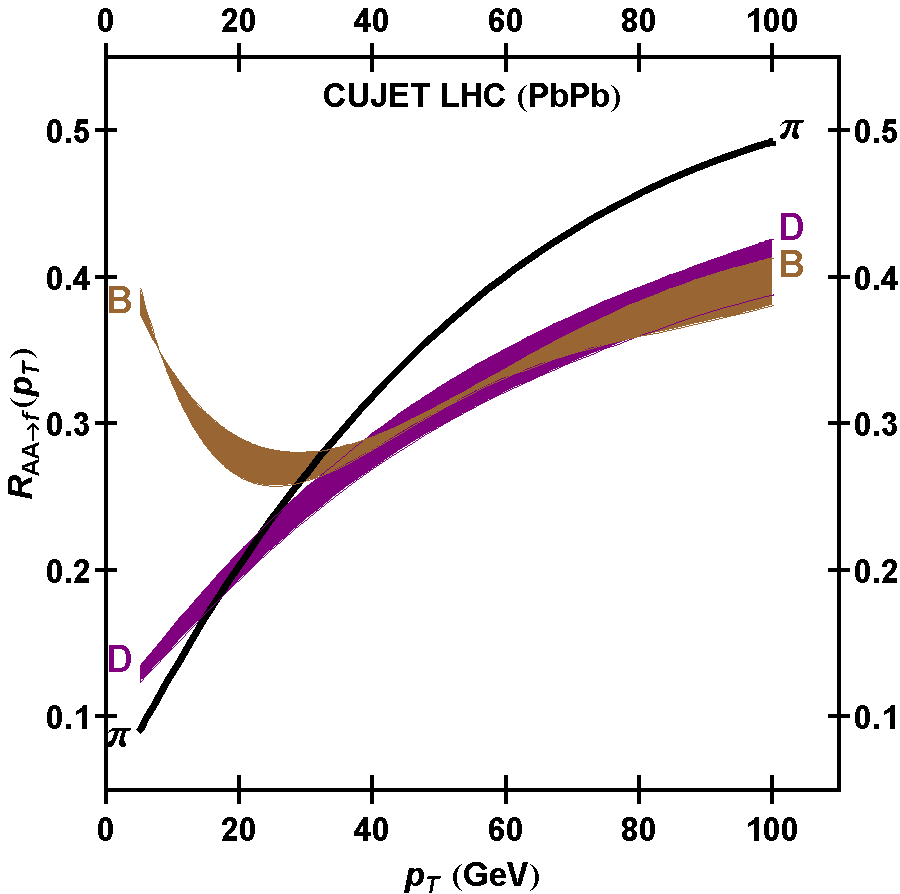}
\caption{Illustration of jet flavor tomography level crossing pattern
  of nuclear modification factors versus $p_T$ at $y=0$ for
  $\pi,D,B,e$ fragmentation from quenched $g,u,c,b$ jets in Au+Au 5\%
  at RHIC (left side) and extrapolated to Pb+Pb 5\% at LHC (right
  side) computed with the dynamic CUJET1.0 model at leading $N=1$
  order in opacity.  The opacity is constrained at RHIC, given
  $dN/dy(RHIC)=1000$, by a fit to a reference
  point $R_{AuAu}^\pi(p_T=10\;{\rm GeV})=0.2$ setting $\alpha_s=0.3$.
  The extrapolation to LHC assumes $dN_{ch}/d\eta$ scaling of the
  opacity as measured by ALICE \cite{ALICEdNdy}.  The
  $D,B,e$ bands reflect the uncertainty due to the choice of NLO or
  FONLL initial production spectra.
  Note the possible inversion of $\pi,D,B$ levels predicted by CUJET
  at high $p_T$ at LHC and a partial inversion at RHIC arising from
  competing dependences on the parton mass of energy loss and
  of initial pQCD spectral shapes.  }
\end{figure*}

Jet quenching observables such as $R_{AA}$ can provide quantitative tomographic
information about the QGP density evolution and the dynamics between the jet and
the medium. In order to provide more stringent constraints on the energy loss
models and at the same time to better discriminate among them, it is necessary
to perform simultaneous constrained predictions of as many observables as
possible. In this study, even though we will be focusing on the nuclear
modification factor only, we will examine in detail its dependence on the center
of mass energy of the collision (which in turn regulates the
density and temperature of the plasma) and its dependence on the jet parton
flavor. Other fundamental dependencies, such as rapidity and centrality, will be
object of future studies.

In CUJET the strong coupling constant $\alpha_s$ is the only free parameter that
we fit to the data: constraining one reference $p_T=10$ GeV point of pion
$R_{AA}^\pi=0.2$ at RHIC, we set $\alpha_s=0.3$. The extrapolation to LHC is
then parameter free, assuming that $\alpha_s$ does not vary with $\sqrt{s}$. The
results are shown in Fig.1.
The most striking feature is the novel inversion of the $\pi<D<B$ $R_{AA}$
hierarchy ordering at high $p_T$, in spite of the unmodified hierarchy of the
parton relative energy loss $(\Delta E /E)_u \approx (\Delta E /E)_c \ll (\Delta
E /E)_b$. This effect is mostly due to the steeper initial invariant jet
distributions of c and b jets at RHIC \cite{Vogt}.
The same novel feature, on a broader energy range, can be observed at LHC
(figure on the right).
It becomes then evident the importance of experimentally isolating and observing
the separate contribution of D and B mesons, since the mass splitting between c
and b jets is a particularly robust prediction of pQCD in a deconfined QCD
medium.

\begin{figure*}[tbh]
\label{Fig2}
\includegraphics[width=3.5in
]{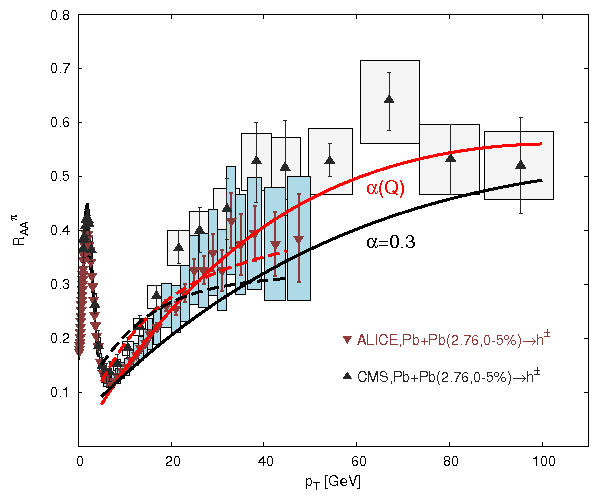}
\hspace{0.25in}
\includegraphics[width=2.5in,
]{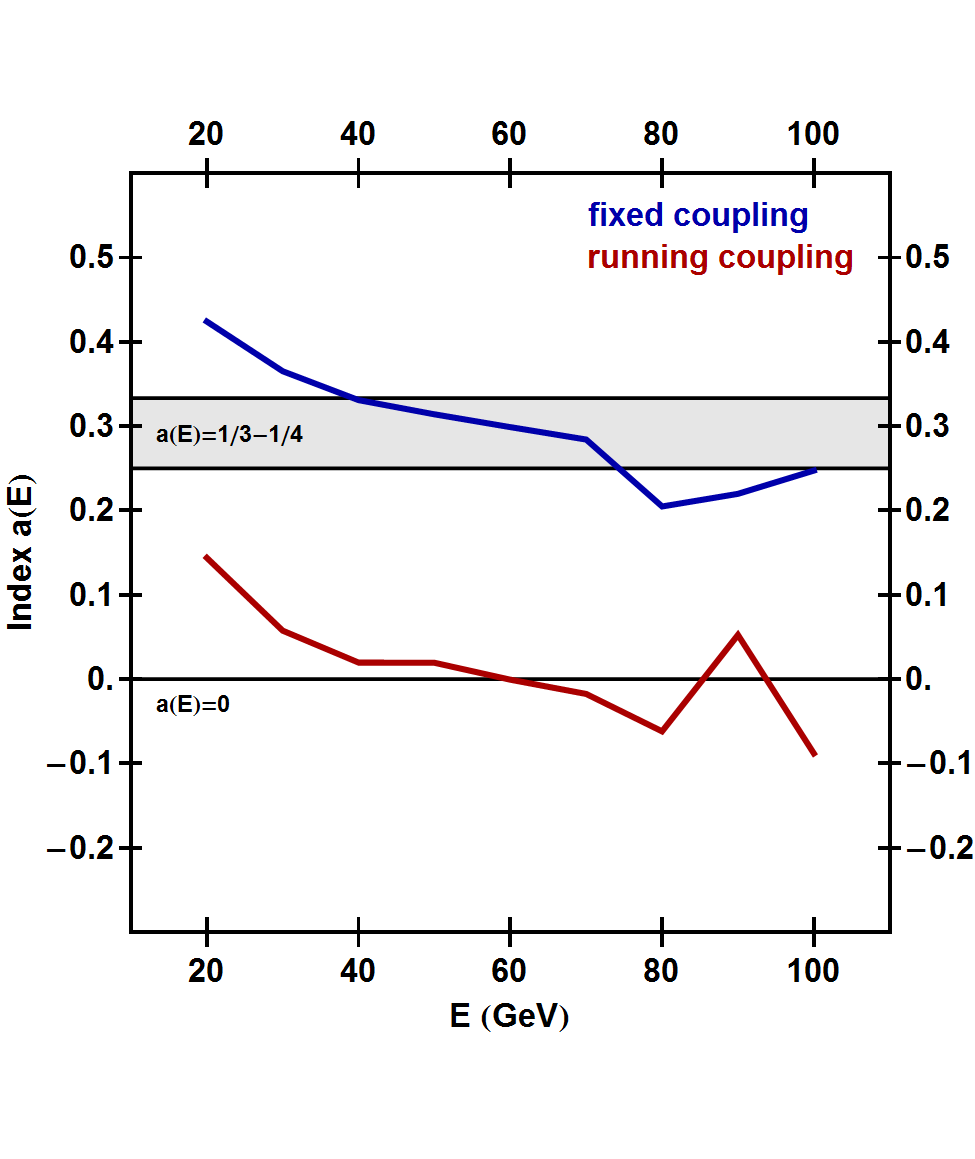}
\caption{(Left) Pion $R_{AA}$ at LHC (solid lines) and RHIC (dashed lines). In black the fixed coupling CUJET results, constrained at $p_T=10$ GeV RHIC with $\alpha_s=0.3$. In red the running coupling CUJET results, constrained
at $p_T=40$ GeV LHC with $\alpha_0=0.4$. Notice the unmodified behavior of RHIC $R_{AA}$ in the range of energies $5-20$ GeV, which makes the running coupling extrapolation to RHIC compatible with present data \cite{PHENIX}. Central $0\%-5\%$ preliminary ALICE and CMS $h\pm$
LHC data \cite{LHClatest} (brown and gray triangles, respectively) are compared to predictions. (Right) Illustration of the
power law
index $a(E)$ as a function of jet energy extracted from a CUJET Bjorken
expanding $5$ fm brick-like simulation of light quark quenching (radiative
only). A three point moving average is shown in the figure.}
\end{figure*}

Although both the pion and non-photonic electron results now agree within the
experimental error to the RHIC data, possibly settling the issue of the ``heavy
quark puzzle'', the LHC extrapolation seems to
systematically overpredict the quenching suffered by the jets in the plasma
\cite{WHMG}.
Furthermore, more recent preliminary results published by the ALICE and CMS
collaborations \cite{LHClatest} appear to indicate a steeper rise of $R_{AA}$ in
the range of momenta of interest ($30-100$ GeV), suggesting the possibility of reduced
coupling observed at LHC energies and densities \cite{Betz}.
Motivated by these findings, we relaxed the effective fixed alpha approximation
and utilized a one-loop order running coupling, parametrized as follows
\cite{Zakharov_RunningCoupling}:
\be
\alpha_s(Q^2)= \left\{ \begin{matrix}
\alpha_0\equiv\frac{2\pi}{9\ln(Q_0/\Lambda_{QCD})} \; (Q\leq Q_0) \; ;\\
\frac{2\pi}{9\ln(Q/\Lambda_{QCD})} \;(Q>Q_0) \; .\\
\end{matrix}\right.
\ee
Again we choose to keep $\alpha_0$ as the only free parameter of the model. The
choice of scale $Q$, of the order of $1$ GeV, is somewhat arbitrary.
To address this systematic source of uncertainty, we let it vary while fixing
the parameter $\alpha_0$ to fit one chosen pion $R_{AA}^{LHC}(p_T=40GeV)=0.35$ point.
We include running coupling effects in both the radiative and elastic
\cite{PeignePeshier}
contribution to the total energy loss.
The results are shown in Fig.2.

Observing the figure on the left, it is evident that the overall shape of
$R_{AA}$ across the broad range of $p_T$ under consideration is changed with respect to the
previous fixed coupling results. Besides appreciating the more satisfactory
agreement with
data, both at LHC and RHIC (in the latter case our predictions are almost left
unchanged given the restricted range of energies at play), it is surprising to
note how the effective energy dependence itself of the energy loss appears to be
modified (figure on the right).
Assuming in fact a simplified model for the energy loss
\be
\frac{\delta E}{E}=\kappa E^{a-1} L^b \rho^c \;
\ee
and extracting the index $a(E)$ from our results, it seems that the pQCD $\ln E
\approx E^{1/3}-E^{1/4}$ characteristic LPM dependence of the energy loss is
canceled when the running coupling effects are included.

\section{Conclusions}
The CUJET model has been applied to study the flavor and $\sqrt{s}$ dependence of the nuclear modification factors for central collisions at mid-rapidity. With one free parameter ($\alpha_s$) used to fit the pion data at RHIC, we have predicted a novel level crossing pattern of $R_{AA}$ for different flavors. The inclusion of running coupling effects in the model has instead improved the agreement with the recent LHC data without affecting the RHIC sector in the range of energies currently probed. Further data may give additional indication of running effects and lead to stronger conclusions with respect to the energy dependence of the energy loss itself.

{Acknowledgments}: We acknowledge support by US-DOE Nuclear Science Grant No.
DE-FG02-93ER40764.

\bibliographystyle{elsarticle-num}

\end{document}